\documentclass[final,3p,times,twocolumn,authoryear,longtitle]{elsarticle}

\usepackage{amssymb}
\usepackage{amsmath}

\usepackage{booktabs,caption}
\usepackage{threeparttable} 
\usepackage[T1]{fontenc}
\usepackage[utf8]{inputenc}
\usepackage{color} 
\usepackage{hyperref}

\usepackage{aas_macros}

\newcommand{\td}[1]{\, \mathrm{d} #1 \,}

\newcommand{\g}{\ensuremath{\gamma}}

\newcommand{\E}[1]{\times 10^{#1}}

\newcommand{\fermi}{\textit{Fermi}-LAT}
\newcommand{\fvar}{\ensuremath{F_\text{var}}}
\journal{Journal of High Energy Astrophysics}

\begin{document}

\begin{frontmatter}



\title{20 years of monitoring: PKS\,2155$-$304 and PKS\,1510$-$089 in the eyes of Swift and Fermi.\\ I. The case of PKS\,2155$-$304} 

\author[ifj]{Alicja Wierzcholska}
\ead{alicja.wierzcholska@ifj.edu.pl}
\author[lsw,nwu]{Michael Zacharias}
\ead{m.zacharias@lsw.uni-heidelberg.de}
\affiliation[ifj]{organization={Institute of Nuclear Physics, Polish Academy of Sciences},
            addressline={ul. Radzikowskiego 152},
            postcode={31-342},
            postcodesep={},
            city={Krak\'{o}w},
            country={Poland}}
\affiliation[lsw]{organization={Landessternwarte, Universit\"{a}t Heidelberg},
            addressline={K\"{o}nigstuhl 12},
            postcode={D-69117},
            postcodesep={},
            city={Heidelberg},
            country={Germany}}
\affiliation[nwu]{organization={Centre for Space Research, North-West University},
            city={Potchefstroom},
            postcodesep={},
            postcode={2520},
            country={South Africa}}

\begin{abstract}
We present a comprehensive 20-year multiwavelength variability study of the blazar PKS\,2155$-$304, one of the most luminous and extensively monitored high-frequency-peaked BL~Lac objects in the southern hemisphere.
Using \textit{Fermi}-LAT $\gamma$-ray data together with \textit{Swift}-XRT and UVOT observations spanning 2005–2024, we trace the long-term evolution of its flux, interband correlations, and spectral behaviour across the optical, X-ray, and $\gamma$-ray bands.
All flux distributions are compatible with log-normality. Interestingly, the optical domain exhibited a notable baseline change around 2009, but this has no strong influence on the fit of the flux distribution. 
While interband flux–flux correlations are found, no stable temporal lags emerge. This implies varying correlation patterns between epochs.
The X-ray emission displays a robust harder-when-brighter trend, however with epoch-dependent slopes, while the $\gamma$-ray spectra show only mild flux dependence. The fractional variability increases systematically with energy within a given radiation component. 
No direct correlation of the year-wise fractional variability with the corresponding average flux could be found.
Interestingly, a pronounced X-ray spectral upturn, detected during a low state in 2012, points to an additional radiative component. As the connection from this upturn to the \g-ray spectrum is not smooth, it probably is not the onset of the inverse-Compton component, but more likely points either to a hadronic contribution or an additional spatially-separate emission zone.
These findings reveal the complexity of variability patterns in PKS\,2155$-$304 and the non-uniform nature of its particle acceleration and emission processes.
\end{abstract}




\end{frontmatter}



\section{Introduction} \label{sec:intro}
Blazars are a class of active galactic nuclei (AGN) characterised with  polarised and strongly variable non-thermal continuum emission.
Within the framework of the unified model \citep{Urry}, blazars are understood as AGN whose relativistic jets are oriented at small angles with respect to the observer’s line of sight.

Blazars are commonly divided into two main types: BL Lacertae (BL Lac) objects and flat-spectrum radio quasars (FSRQs). 
This classification has been based on the equivalent width of their optical emission lines. 
The broadband emission of blazars spans the entire electromagnetic spectrum, from radio frequencies to very-high-energy (VHE, $E>100\,$GeV) $\gamma$ rays \citep[e.g.][]{Wagner09,hess07pks}.
A characteristic feature of these sources is variability across all energy bands, occurring on timescales ranging from minutes to years \citep[e.g.][]{hess07pks, Wierzcholska15}.
Such flux variations are frequently accompanied by spectral changes \citep[e.g.][]{Xue,  2016MNRAS.458...56W}.
 
The blazar PKS\,2155$-$304, located at a redshift of $z=0.117$ and classified as a high-frequency peaked BL Lac object (HBL), is among the brightest blazars in the southern hemisphere. 
Initially detected in the radio band during the Parkes survey \citep{Shimmins}, it was later identified as a BL Lac-type source by \citep{Hewitt}.
Since its discovery, PKS\,2155$-$304 has been extensively monitored in all energy regimes. 

As one of the brightest known BL Lac objects, PKS\,2155$-$304 has been the subject of numerous detailed optical and UV observational campaigns \citep[e.g.,][]{Courvoisier, Pesce, Pian}.
These campaigns confirmed the intraday variability of the blazar in the optical range, with the shortest recorded flux-doubling timescales of roughly $15\,$min, although such rapid changes were observed only during a few isolated epochs \citep[e.g.,][]{Paltani, Heidt}.
The optical variability exhibits also a red-noise character, persisting over timescales of up to three years \citep{Kastendieck}.
In several epochs, the source displayed a clear bluer-when-brighter trend \citep[e.g.,][]{Paltani, ATOM_monitoring}, but this trend is not present in the long-term datasets. 

Studies  of the optical-UV and X-ray temporal and spectral properties of PKS\,2155$-$304 show that the peak of the synchrotron component of the spectral energy distribution (SED) typically lies around the UV range, only occasionally shifting toward optical wavelengths \citep[e.g.,][]{Foschini}.
The variability amplitudes in the optical and UV bands are consistently much smaller than those observed in X-ray range. 

The X-ray variability of the source follows a red-noise pattern, with a characteristic timescale on the order of days and minimal power at intraday timescales \citep{Zhang99, Zhang02, Tanihata, Kataoka01}.
The variability amplitude is found to correlate with both the source flux and photon energy. Spectral changes in the X-ray band are usually consistent with a  “harder-when-brighter” trend \citep[e.g.,][]{Zhang05, Zhang06, Abramowski12}.

High-energy (HE, $E>100\,$MeV) $\gamma$-ray emission from PKS~2155$-$304 was first detected with EGRET in the $30\,$MeV to $10\,$GeV energy range \citep{Vestrand}, while its VHE $\gamma$-ray emission was discovered with the University of Durham Mark~6 telescope \citep{Chadwick}.
The blazar was frequently observed with the High Energy Stereoscopic System (H.E.S.S.) since 2002 \citep{Aharonian2005}.
From 2004 onward, PKS~2155$-$304 has been monitored with the four-telescope H.E.S.S. array, leading to the discovery of two extraordinary $\gamma$-ray flares in July 2006.
These flares reached peak fluxes about 40 times higher than the average and exhibited VHE flux and revealed doubling timescales of only a few minutes \citep{hess07pks, Aharonian2009_flare2}.
The multiwavelength (MWL) observations collected around the second flare revealed a strong X-ray--VHE correlation at high flux levels, which weakened at lower fluxes.
The X-ray flux variations (by a factor of $\sim 2$) were notably smaller than those observed in the VHE band, and no consistent optical--VHE correlation was found during the flaring episodes \citet{Abramowski12}.  

A detailed statistical analysis of H.E.S.S. observations from 2005--2007 \citep{Abramowski2010} confirmed that the VHE variability follows a red-noise pattern, with relatively short characteristic timescales ($\lesssim 1$\,d), especially during the 2006 flares \citep[see also][]{hess07pks}.
The variability amplitude, quantified by the fractional r.m.s., was correlated with photon energy, while the excess r.m.s. scaled with flux. Pronounced spectral variability in the VHE band was found, with distinct patterns between low-and high-flux states.
\cite{HESS_ATOM} reported a lack of any universal relation between VHE $\gamma$-ray and optical fluxes based on 2007-2009 observations of the source.

The studies of the HE and VHE $\gamma$-ray emission from this BL~Lac object over multi-year timescales during its quiescent state reveal a log-normal flux distribution. The power-spectral density (PSD) is consistent with flicker noise (PSD index $\beta_{\rm PSD} \approx 1.1$) in the VHE band, while the HE data yield a similar index ($\beta_{\rm PSD} \approx 1.2$), suggesting scale-invariant variability across bands \citep{2155_lognormal}.
The 2013 to 2016 MWL monitoring of the blazar from the optical to the VHE $\gamma$-ray range \citep{2155_monit2015} disentangles a diverse set of variability behaviours including a long-lasting “orphan” optical/UV flare, indications of correlation between X-ray and VHE $\gamma$-ray fluxes,  but  also a lack of universal optical–X-ray or optical–$\gamma$-ray correlations.

The goal of our paper is to analyse the evolution of the long-term MWL flux and spectral variations from the optical domain to \g\ rays. We employ 20 years of data from PKS\,2155$-$304 ranging from 2005 to 2024 taken with the \textit{Swift} and \textit{Fermi} observatories. We study the interband correlations, the spectral characteristics, flux distributions, as well as variability aspects of the various bands.

The paper is organised as follows. In Sec.~\ref{sec:data}, the data analysis is presented, while Sec.~\ref{sec:varia} is devoted to the temporal variations of fluxes and spectra. In Sec.~\ref{sec:xupturn}, we present the curious case of a convex X-ray spectrum. We discuss and summarise our findings in Secs.~\ref{sec:discus} and~\ref{sec:sum}, respectively.
A follow-up study \citep{ZW26} will extend this work by analysing the flat-spectrum radio quasar PKS\,1510$-$089 including a comparison of both blazars.

%
%
\section{Data analysis} \label{sec:data}

\subsection{Fermi-LAT}
The HE \g-ray observations of PKS\,2155$-$304 collected with the Large Area Telescope (LAT) aboard the \textit{Fermi} Gamma-ray Space Telescope  cover the period from the mission start at August 04, 2008, to August 08, 2024, corresponding to 16 years of observations of the blazar.
The analysis was performed using \textit{Fermi} Science Tools version 2.2.0, fermipy version 1.2.2  and instrument response functions \verb|P8R3_SOURCE_V3|.
The events are selected within a $15^\circ$ radius of the source position, applying standard quality cuts including a zenith angle $<90^\circ$ and the \verb|DATA_QUAL|~$> 0$ filter to exclude periods of poor data quality.
The energy range was restricted to $100\,$MeV - $500\,$GeV. 
The background model included all 4FGL-DR3 catalog sources within $12^\circ$ of the target, with sources within $3^\circ$ having free spectral parameters during the likelihood fitting procedure.
Galactic and isotropic diffuse emission components were modelled using the standard templates \verb|gll_iem_v07.fits| and \verb|iso_P8R3_SOURCE_V3_v1.txt|, respectively.
Source flux and spectral parameters were determined using a binned likelihood analysis with 10 energy bins per decade and spatial bins of $0.1^\circ$.
Eight additional  point-like sources:  J2211.9-2923, J2151.5-2743, J2212.6-2654, J2217.7-2737, J2134.6-2705, J2146.2-3606, J2139.3-2443 and J2128.8-3510 were added to the background model.
They were identified by the method \verb|find_sources| from the fermipy package, requiring a test statistic $(\mathrm{TS}) > 5$.

The data are separated into intervals of 4\,d duration.
The spectral reconstruction per time bin is done using a power-law model,

\begin{align}
    \frac{\td{F}}{\td{E}} = N_0 \left( \frac{E}{E_0} \right)^{-\Gamma}
    \label{eq:powerlaw},
\end{align}
with normalisation $N_0$ at fixed reference energy $E_0= 1\,$GeV, and the spectral index $\Gamma$. We also considered a fit with a log-parabola model,

\begin{align}
    \frac{\td{F}}{\td{E}} = N_0 \left( \frac{E}{E_0} \right)^{-\alpha-\beta\log{(E/E_0)}}
    \label{eq:logparabola},
\end{align}
with the spectral index $\alpha$, the curvature $\beta$, and the $\log$ denoting the natural logarithm. However, the number of flux bins, where the log-parabola model would significantly improve the fit is less than 2\%. Hence, we use the power-law reconstruction for all bins.

In the following, only light curve bins fulfilling these criteria are used: $TS>9$, and $F/\Delta F >1.0$, where $F$ and $\Delta F$ are the integrated flux of a time bin and its uncertainty, respectively. These criteria remove insignificant flux bins. 
The resulting light curve is shown in Fig.~\ref{fig:2155_mwl_lc}(a), while the evolution of the spectral index is shown in panel (b) of the same figure. The weighted average\footnote{The weights are determined from the statistical error of the observed quantity.} flux is $(9.0\pm 0.1)\E{-8}\,$ph/cm$^2$/s, while the average index is $1.796\pm 0.005$. The flux is strongly variable, while the spectral index is compatible with a constant.

\subsection{Swift-XRT and Swift-UVOT}

Observations carried out between 2005 and 2024, corresponding to ObsIDs 00035027001-00033300085, were analysed using the HEASoft software package v6.35.1. All event files were cleaned and calibrated with the \verb|xrtpipeline| task. Data from both Photon Counting (PC) and Windowed Timing (WT) modes within the 0.3–10\,keV energy range were included in the analysis. For spectral fitting, the data were binned using the \verb|grppha| tool to ensure a minimum of 20 counts per bin. The spectra were modelled with a power-law, Eq.~\eqref{eq:powerlaw}, and a log-parabola component, Eq.~\eqref{eq:logparabola}, and a fixed Galactic hydrogen column density of $1.28 \times 10^{20}$\,cm$^{-2}$ \citep{HI4PI}, using the \verb|XSPEC| software \citep{Arnaud96}. In all fits, N$_H$ was kept frozen. The fits were compared using the F-test \citep[e.g.][]{Bevington}. A preference (p-value $<10^{-2}$) for the curved model was obtained in around 90\% of the cases, which is why the log-parabola model is kept for all XRT observations.



\begin{figure*}
\centering
\includegraphics[width=0.98\textwidth]{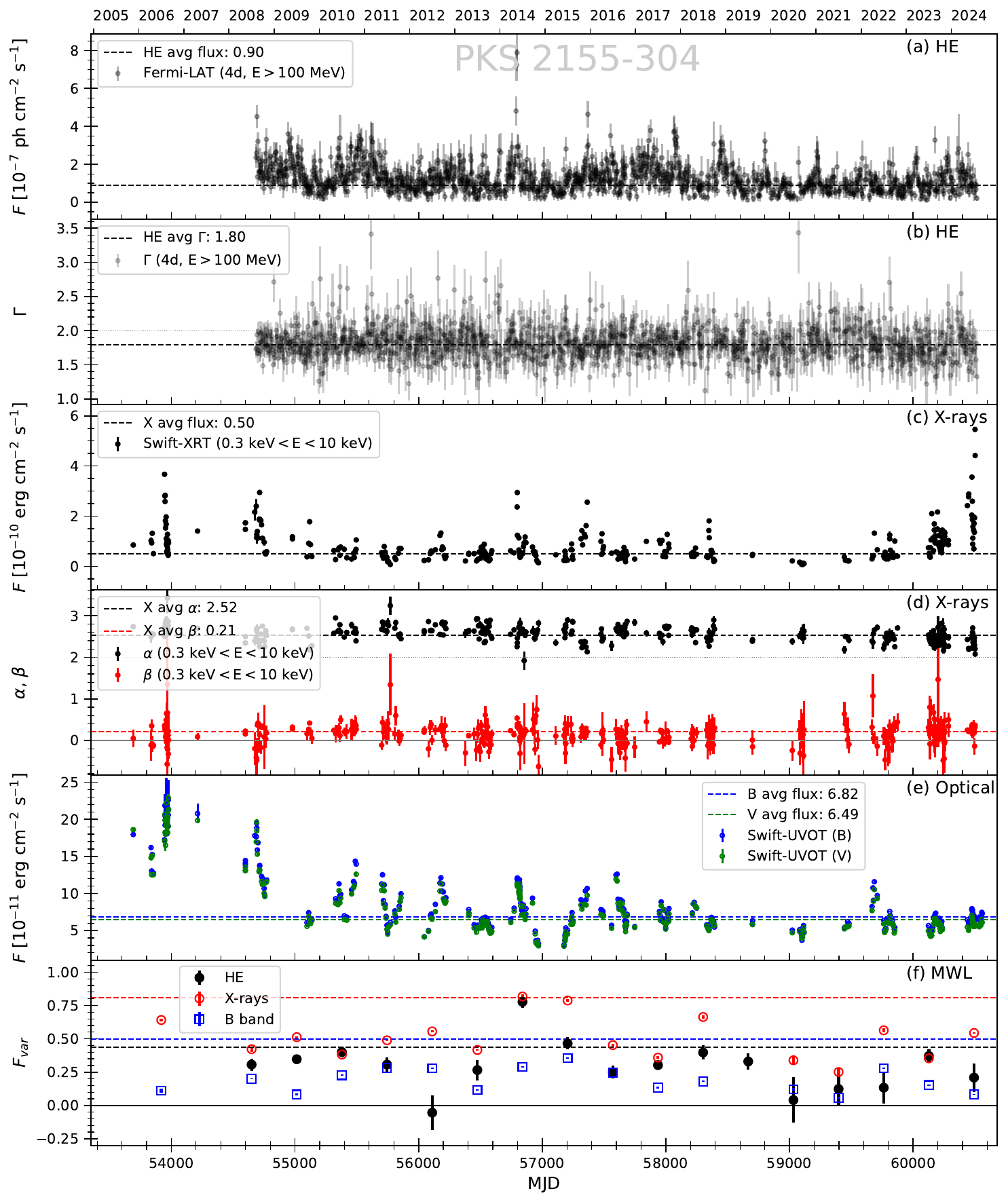}
\caption{Time-evolution of PKS\,2155$-$304 from 2005 to 2024. 
(a) HE \g-ray light curve from \fermi. The black dotted line marks the average flux of the data shown. 
(b) HE \g-ray spectral index with the black dashed line indicating its weighted average. The grey dotted lines marks a value of $2.0$.
(c) Same as (a) but for the X-ray data from \textit{Swift}-XRT.
(d) Same as (b) but for the X-ray data from \textit{Swift}-XRT.
(e) B and V band light curves from \textit{Swift}-UVOT with dashed lines indicating the respective average.
(f) Year-wise \fvar\ values for the \g-ray, X-ray and B bands. The coloured dashed lines indicate the respective \fvar\ values considering all data of a given band. 
}
\label{fig:2155_mwl_lc}
\end{figure*}

The X-ray light curve is plotted in Fig.~\ref{fig:2155_mwl_lc}(c) with the spectral parameter evolution presented in panel (d). The weighted average of the flux is $(5.02\pm0.01)\E{-11}\,$erg/cm$^{2}$/s, while it is $2.524\pm0.002$ and $0.211\pm0.005$ for the index and curvature, respectively. All X-ray quantities are significantly variable.
Much like with \fermi, many individual observations are compatible with a simple power law, but we stick to the log-parabola models for consistency.

Simultaneously, the source was observed with the UVOT instrument onboard \textit{Swift} using the U (345\,nm), B (439\,nm), and V (544\,nm) filters. For each observation corresponding to the aforementioned ObsIDs, instrumental magnitudes were computed using \verb|uvotsource|, extracting photons from a circular region with a 5'' radius. Background estimation was performed using a nearby, source-free circular region of 10'' radius. Flux conversion factors were adopted from \cite{Poole08}. All UVOT data were corrected for Galactic extinction using a reddening value of $E(B-V) = 0.0185$\,mag \citep{Schlafly}. The extinction corrections for each filter were applied using the $A_{\lambda}/E(B-V)$ ratios from \citep{Giommi06}.
In addition, the data used in the following analysis were corrected for the host galaxy contribution, adopting the elliptical galaxy template of \cite{Fukugita} together with Gunn i-band observations from \cite{Falomo}, assuming a de~Vaucouleurs profile for the starlight.

The B and V band light curves are shown in Fig.~\ref{fig:2155_mwl_lc}(e). The average fluxes are $(6.69\pm0.01)\E{-11}\,$erg/cm$^2$/s and $(6.19\pm0.01)\E{-11}\,$erg/cm$^2$/s for the B and V band, respectively. Both light curves are strongly variable. As the two bands are directly correlated with each other, we will mostly use the B band in the following.

\section{Variability study} \label{sec:varia}

In this section, we study the details of the temporal variations within and between the bands. This includes analyses of correlations, flux distributions, and the fractional variability parameter.

\subsection{Correlations} \label{sec:corrs}
\begin{figure*}
\centering
\includegraphics[width=0.98\textwidth]{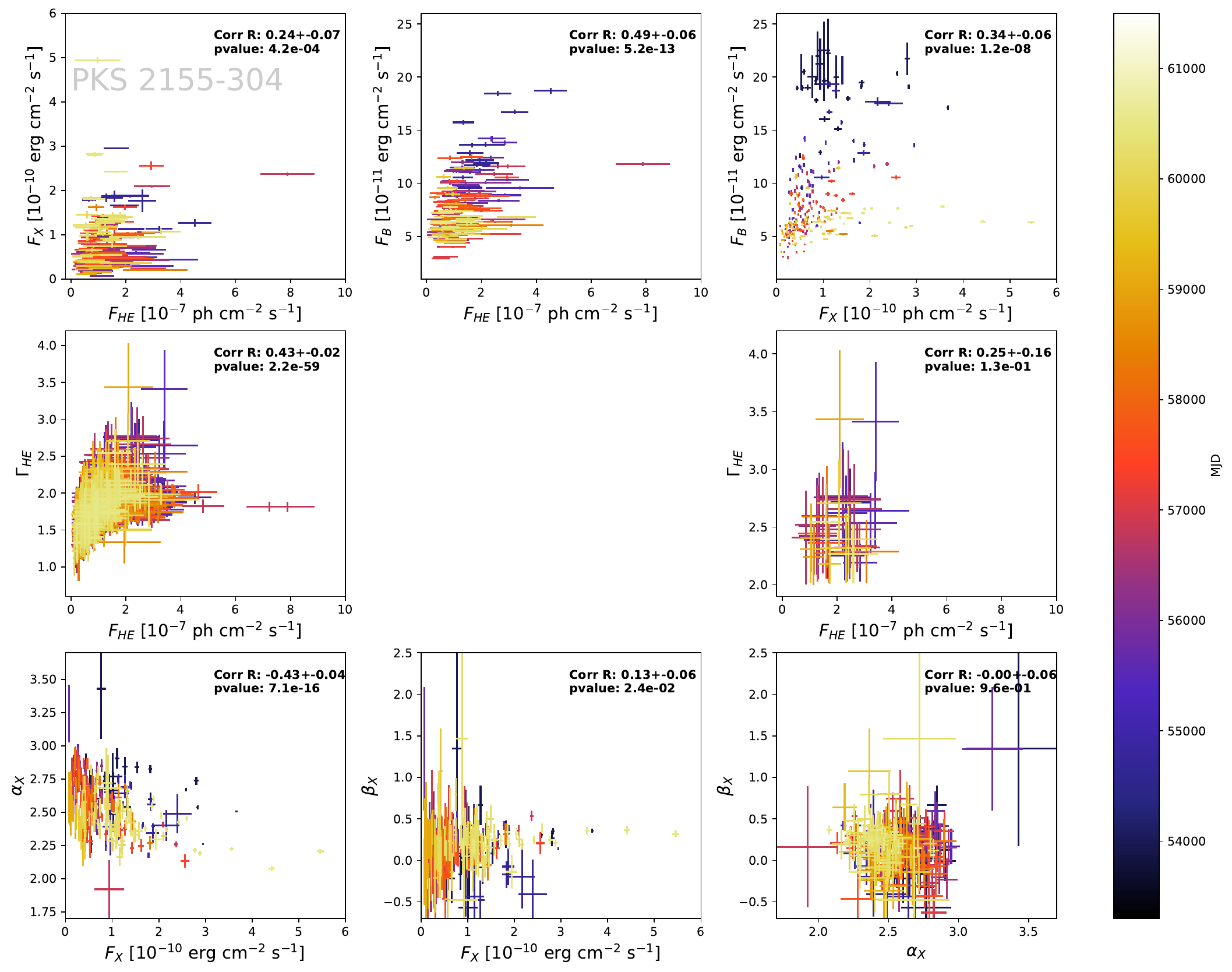}
\caption{(Top row) Correlation plots of flux vs flux for X-rays vs HE \g\ rays (left), B band vs HE \g\ rays (middle), and B band vs X-rays (right).
(2nd row) HE \g-ray index vs flux (left), and the same but only for soft spectra (right). The individual points of the latter are given in Tab.~\ref{tab:2155_HE_soft_spectra}.
(Bottom row) X-ray index vs flux (left), curvature vs flux (middle), and curvature vs index (right).
In each panel, the Person R coefficient for the correlation is given, as well as the p-value of the probability that a similar or higher R coefficient could be obtained from an uncorrelated system.
}
\label{fig:2155_flux_flux}
\end{figure*}

\begin{figure*}
\centering
\includegraphics[width=0.98\textwidth]{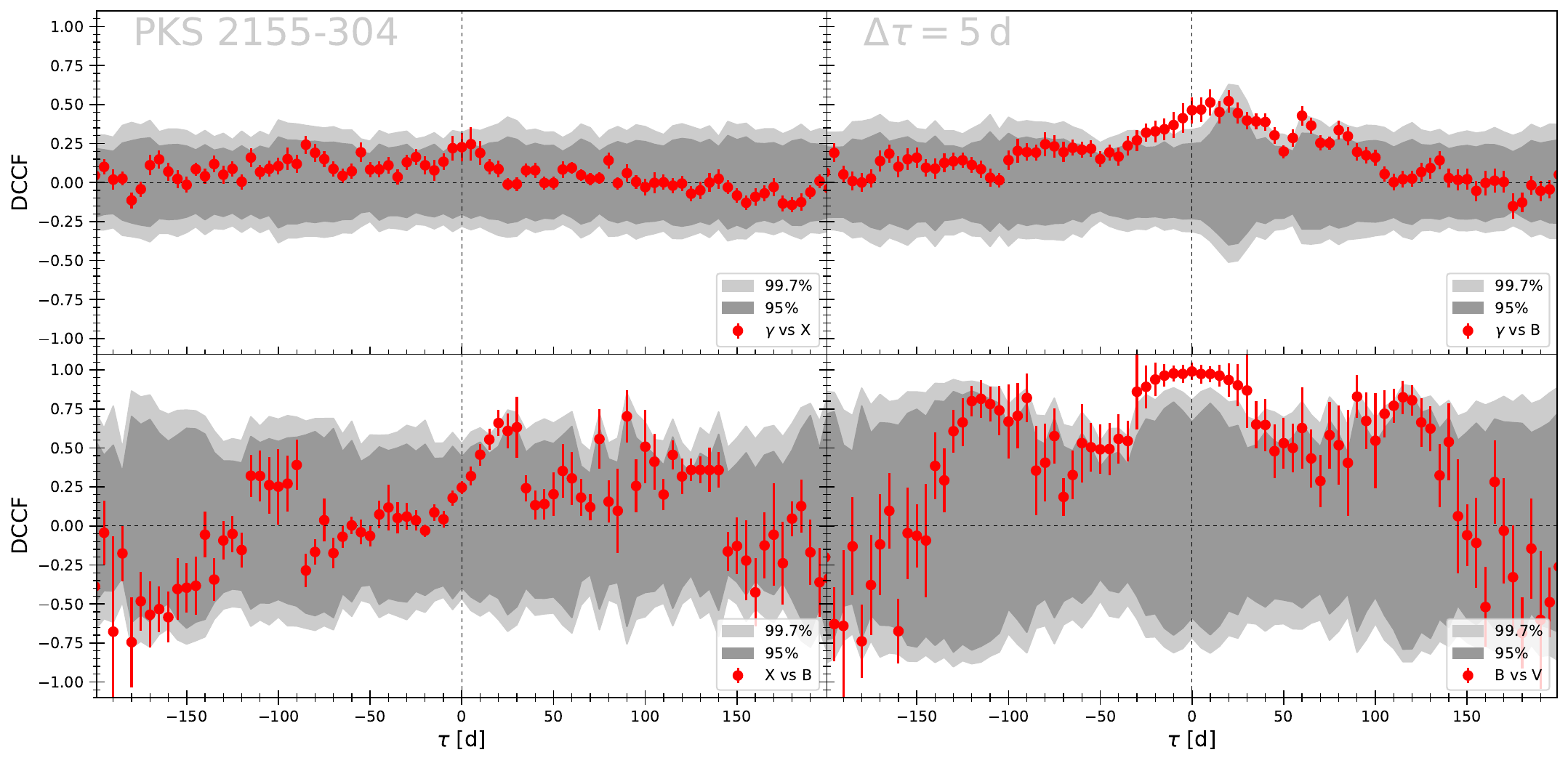}
\caption{DCCF with a time resolution of $\Delta\tau=5\,$d of PKS\,2155$-$304 between the various light curve bands as labelled. The grey bands mark the confidence intervals as indicated.
}
\label{fig:2155_dccf}
\end{figure*}

\begin{figure}
\centering
\includegraphics[width=0.48\textwidth]{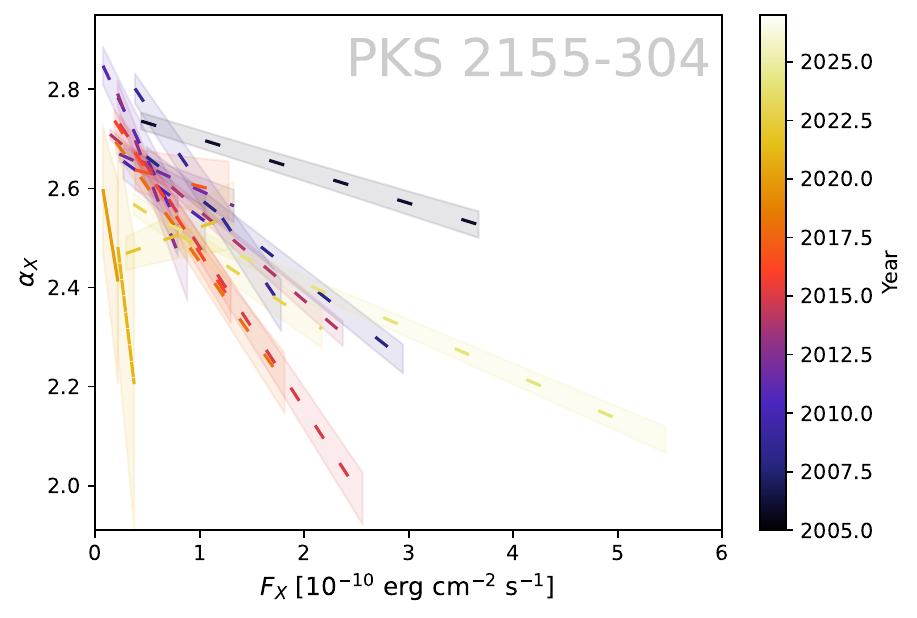}
\caption{X-ray spectral index versus integrated X-ray flux for PKS\,2155$-$304. Dashed lines show  year-wise (years indicated by colour) weighted linear fits to index-vs-flux data, while the shaded regions are constructed from the errors on the fit parameters. The gaps between dashes indicate the goodness-of-fit with large gaps representing a worse fit or higher rejection significance.
}
\label{fig:2155_xind}
\end{figure}

Plotting the fluxes of various bands against each other gives an immediate impression on direct correlations between the respective bands. This is shown in the top row of Fig.~\ref{fig:2155_flux_flux}. Given the different integration times of observations for the 3 bands, the flux points have been matched and combined as follows. For the X-ray versus \g-ray correlation, all X-ray observations within a 4\,d \g-ray bin have been averaged to a single value. The same was done with the B band data for its correlation with the \g-ray data. For the B band versus X-ray correlation, all observations taken on the same day (i.e., the same integer MJD) have been averaged, so that there is only one point per observation day. As some \fermi\ light curve bins are discarded (see Sec.~\ref{sec:data} for the selection criteria), not all X-ray or B band observations have a \g-ray counterpart. 

Evidently from the top row of Fig.~\ref{fig:2155_flux_flux}, there is some form of positive correlation between the bands. This is also supported by the Pearson R coefficient, which is given in each panel. We also provide the corresponding p-value, which indicates the probability that the R coefficient can be reached or surpassed from an uncorrelated system. If we take $p=0.5\%$ as the significance threshold, we find that all three correlations are indeed significant. However, no obvious tracks emerge indicating that the indicated positive correlations differ from event to event. Strikingly, the X-ray flare in 2024 has little to no conterparts in the optical and \g-ray bands.

We further studied the interband correlations employing the discrete cross-correlation function \citep[DCCF,][]{EdelsonKrolik1988}. The DCCF of time difference $\tau$ is defined as

\begin{align}
    \text{DCCF}(\tau) = \frac{1}{N} \sum_{i,j} \frac{(a_i - \bar{a}) (b_i - \bar{b})}{\sigma_a \sigma_b},
    \label{eq:DCCF}
\end{align}
where $a$ and $b$ are the fluxes of the light curves, with mean values $\bar{a}$ and $\bar{b}$, and standard deviations $\sigma_a$ and $\sigma_b$, respectively. $N$ is the number of pairs $i,j$ in the time interval $\tau$. In order to assess the significances of the correlation peaks, we employed a Monte Carlo method producing 10.000 synthetic light curves based on the power spectrum of light curve $b$ \citep[a detailed description of the procedure is provided in][]{Taylor+26}.

The DCCFs are plotted in Fig.~\ref{fig:2155_dccf}. Of the 3 bands used in Fig.~\ref{fig:2155_flux_flux}, only the DCCF of \g\ vs B shows a significant ($>3\sigma$) correlation located at $\tau=0\,$d but with only a moderate magnitude around $0.4$. Interestingly, the actual peak of the DCCF curve is at a lag of $\tau\sim20\,$d, however this peak has less than $3\sigma$. An even higher peak at a similar lag can be found in the X vs B curve, but this one is also compatible with the $3\sigma$ band. The \g\ vs X curve has a small peak at around $\tau=0\,$d, but also not significant. The lags between 20 and 30\,d are related to the cadence of observations, which are normally triggered and often related to observations of ground-based observatories that might be influenced by moon light. As the Monte Carlo routine takes care of the observational cadence and related issues, it explains why these peaks are not significant.

We also show the DCCF curve of the two optical bands, B vs V. The light curves shown in Fig.~\ref{fig:2155_mwl_lc} suggest a direct correlation, and the DCCF curve confirms this with a significant DCCF value close to unity at $\tau=0\,$d. This reaffirms our choice of using only the B band for most of the discussion in this paper. In fact, as these bands are so close together in energy space, the DCCF can also be regarded as a near-autocorrelation. This is highlighted by the symmetry of the DCCF with respect to $\tau=0\,$d. 
The broad peak, which is nearly flat within errors until $\tau=30\,$d, is also related to the cadence of the observations. This is highlighted by the Monte Carlo result, as the significance bands show a similar shape. The slow drop towards $\tau\sim70\,$d is a sign of the slow and mostly smooth evolution of the light curves. 

In Fig.~\ref{fig:2155_flux_flux}, we also plot the HE \g-ray and X-ray spectral parameters, $\Gamma$, $\alpha$ and $\beta$, as a function of flux and against each other. In the HE \g\ rays (2nd row, left), the index shows a significant positive correlation with flux (indicative of a softer-when-brighter trend). 
However, it should be pointed out that despite the 4\,d integration for each bin, the fluxes are mostly low. In turn, the significance of the spectral index is also low at the lowest flux states. As the Pearson R coefficient does not consider statistical errors, the significance of the correlations is overestimated\footnote{There is also an apparent tendency for low spectral indices at low fluxes enhancing the apparent correlation.}. Indeed, if one merely derives a weighted fit, it is compatible with the weighted average of the index derived above.

In the right panel of the second row of Fig.~\ref{fig:2155_flux_flux}, we zoom in on HE \g-ray states that are significantly soft. The conditions imposed are (i) $\Gamma-\Delta\Gamma>2$ and (ii) $(\Gamma-\Delta\Gamma)/\Delta\Gamma>3$. The second condition ensures that the spectral parameters are derived with some significance. 
In HBLs, the peak of the high-energy SED component is usually located close to $100\,$GeV or even higher. As a consequence, the spectrum in the HE range is typically hard with an index lower than 2. The implication for HE spectra with an index larger than 2 is that the peak position must have shifted to lower energies, which is unusual for HBL.
The parameters of the resulting points are listed in Tab.~\ref{tab:2155_HE_soft_spectra}. For this subset, the index is not correlated with the flux, while the weighted average of the index is $2.39\pm0.05$. We find 38 such bins, which corresponds to 3\% of the light curve. We do not see any specific clustering nor any correlation with significant MWL behaviour\footnote{While there are two instances, where two subsequent bins are soft, and one case around MJD$\sim$56540, where there are 4 soft bins within roughly 2 months, this does not stick out from the rest. Only 7 bins have corresponding MWL data, which is also in line with the cadence of \textit{Swift} observations.}.

\begin{table}
\caption{Observation date and spectral parameters of the soft HE \g-ray spectra used in Fig.~\ref{fig:2155_flux_flux}, 2nd row (right).}
\label{tab:2155_HE_soft_spectra}
\begin{tabular}{lcc}
\hline
MJD &	$F_{\mathrm{HE}}$	&	$\Gamma_{\mathrm{HE}}$	\\
    & $[10^{-7}\,\text{ph/cm}^2\text{/s}]$ & \\
\hline
54828.7 &	$2.47\pm 0.71$ &	$2.71\pm 0.29$ \\
54880.7 &	$2.30\pm 0.69$ &	$2.25\pm 0.22$ \\
55140.7 &	$1.66\pm 0.86$ &	$2.39\pm 0.34$ \\
55204.7 &	$2.2\pm 1.0$ &	$2.76\pm 0.48$ \\
55348.7 &	$2.84\pm 0.62$ &	$2.19\pm 0.17$ \\
55360.7 &	$2.34\pm 0.81$ &	$2.28\pm 0.25$ \\
55364.7 &	$3.4\pm 1.2$ &	$2.64\pm 0.33$ \\
55424.7 &	$2.34\pm 0.76$ &	$2.62\pm 0.32$ \\
55616.7 &	$3.41\pm 0.85$ &	$3.41\pm 0.52$ \\
55620.7 &	$3.22\pm 0.96$ &	$2.54\pm 0.26$ \\
55904.7 &	$1.31\pm 0.69$ &	$2.43\pm 0.38$ \\
55936.7 &	$1.28\pm 0.68$ &	$2.51\pm 0.42$ \\
56156.7 &	$2.22\pm 0.90$ &	$2.77\pm 0.41$ \\
56288.7 &	$1.35\pm 0.65$ &	$2.60\pm 0.37$ \\
56420.7 &	$2.5\pm 1.1$ &	$2.48\pm 0.33$ \\
56540.7 &	$2.4\pm 1.2$ &	$2.74\pm 0.43$ \\
56620.7 &	$0.86\pm 0.51$ &	$2.41\pm 0.41$ \\
56640.7 &	$1.5\pm 1.0$ &	$2.52\pm 0.42$ \\
56652.7 &	$2.73\pm 0.84$ &	$2.33\pm 0.22$ \\
56664.7 &	$2.6\pm 1.0$ &	$2.66\pm 0.39$ \\
56816.7 &	$2.47\pm 0.69$ &	$2.26\pm 0.22$ \\
57192.7 &	$2.81\pm 0.64$ &	$2.34\pm 0.19$ \\
57388.7 &	$1.29\pm 0.71$ &	$2.45\pm 0.42$ \\
57484.7 &	$1.62\pm 0.68$ &	$2.36\pm 0.31$ \\
58164.7 &	$1.72\pm 0.59$ &	$2.27\pm 0.26$ \\
58180.7 &	$1.63\pm 0.92$ &	$2.59\pm 0.44$ \\
58252.7 &	$1.17\pm 0.33$ &	$2.31\pm 0.22$ \\
58388.7 &	$3.1\pm 1.2$ &	$2.29\pm 0.28$ \\
58832.7 &	$1.15\pm 0.38$ &	$2.24\pm 0.24$ \\
59072.7 &	$2.10\pm 0.89$ &	$3.43\pm 0.60$ \\
59156.7 &	$1.05\pm 0.38$ &	$2.41\pm 0.30$ \\
59364.7 &	$1.03\pm 0.39$ &	$2.31\pm 0.29$ \\
59536.7 &	$1.73\pm 0.42$ &	$2.18\pm 0.18$ \\
59588.7 &	$2.33\pm 0.92$ &	$2.32\pm 0.30$ \\
59808.7 &	$2.08\pm 0.81$ &	$2.71\pm 0.37$ \\
60284.7 &	$2.4\pm 1.0$ &	$2.39\pm 0.29$ \\
60360.7 &	$2.60\pm 0.91$ &	$2.27\pm 0.26$ \\
60388.7 &	$1.60\pm 0.62$ &	$2.54\pm 0.33$ \\
\hline
\end{tabular}
\end{table}

The X-ray parameter correlations are shown in the bottom panel of Fig.~\ref{fig:2155_flux_flux}. The index and flux exhibit a mild, but negative correlation indicative of a harder-when-brighter trend, even though the distribution is relatively broad. There is no correlation between curvature and flux, and neither between curvature and index. Interestingly, a weighted constant fit is significantly rejected (more than $13\sigma$) in all three cases implying significant scatter in the data.

We further study the harder-when-brighter trend in the X-ray band through year-wise weighted linear fits applied to the index-vs-flux data. This is shown in Fig.~\ref{fig:2155_xind}. We require at least 5 data points in a given year to make this fit, which leaves out 2005, 2007, and 2019. All years indicate the harder-when-brighter trend with the sole exception being 2022. Estimating the goodness-of-fit through a $\chi^2$ analysis reveals that only 2020 and 2021 -- the years with the least amount of variability -- are properly fit with a linear line (even though a constant is also compatible with the data). Nevertheless, both the bottom left panel in Fig.~\ref{fig:2155_flux_flux} and Fig.~\ref{fig:2155_xind} suggest that the harder-when-brighter trend is not uniform. That is, the slope of the trend differs from year to year and possibly from flare to flare. The value of the slope appears to be independent of the maximum flux. Ignoring 2020 and 2021, the steepest slope is attained in 2013 with $-0.56\pm0.04$, but the flux in that year does not even reach $1\E{-10}\,$erg/cm$^{2}$/s. Looking only at the years with a minimum maximum flux of $2\E{-10}\,$erg/cm$^{2}$/s, the steepest slope is attained in 2015 with $-0.33\pm0.02$. The least steep slope is exhibited in 2006 with $-0.065\pm0.006$ closely followed by 2024 with $-0.091\pm0.004$. The latter is especially interesting, as it is the year with the highest X-ray flux. It is also worth mentioning that there is no index with a value significantly smaller than $2.0$.
The ``initial index'' of a year (that is the index of the lowest flux value) only plays a minor role, with almost all trends originating between indices of $2.6$ and $2.8$.

\subsection{Flux distribution} \label{sec:fluxdist}

\begin{figure*}[htb]
\centering
\includegraphics[width=0.98\textwidth]{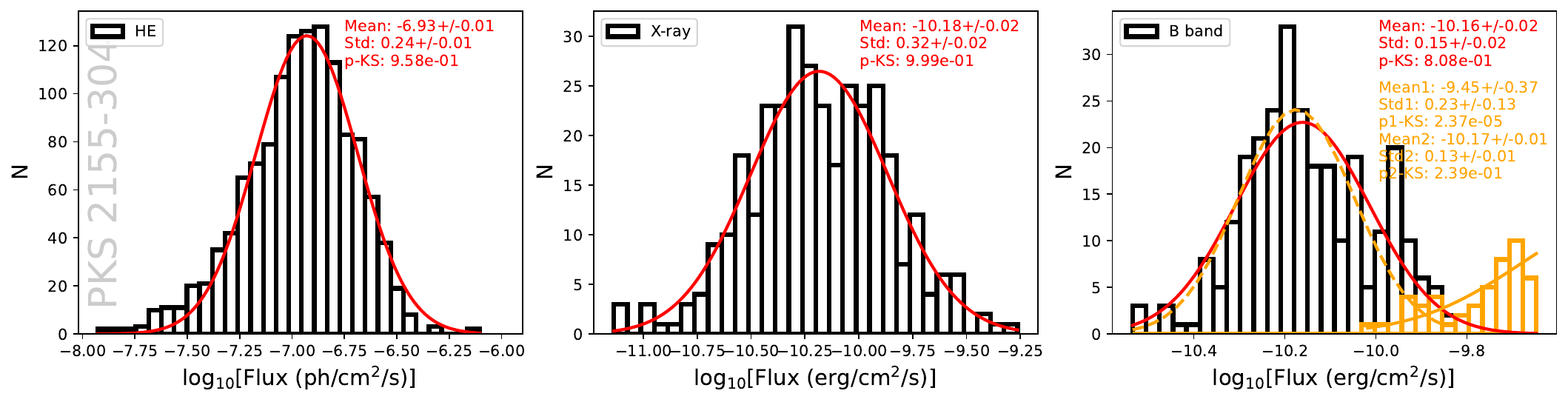}
\caption{Flux distributions of the HE \g-ray (left), X-ray (middle) and B band (right) data using logarithmic binning. The red lines mark Gaussian fits with its mean, standard deviation and the p-value (based on the KS test) of the fit indicated. The golden histogram for the B band is the data taken before 2009. The golden solid line fits the golden histogram (parameters "1"), while the golden dashed line fit the black histogram without the golden data (parameters "2").
}
\label{fig:2155_flux_dist}
\end{figure*}

The distributions of the light curves are shown in Fig.~\ref{fig:2155_flux_dist} with logarithmic binning. Gaussian fits were applied to each distribution. The corresponding fit values, as well as the p-value of the Kolmogorov-Smirnoff (KS) test are given in each panel. All distributions are fitted well implying that the light curves are compatible with being log-normally distributed. The B band distribution is suggestive of an additional peaks at $\sim 2\times 10^{-10}\,$erg/cm$^2$/s. The latter can be understood from the initial higher ``ground-level'' flux in 2005-2008.

We further analysed the optical ground level change by adding a histogram for the B band for the pre-2009 data only (golden histogram) and made separate fits for the pre- and post-2009 histograms (golden solid and dashed curves, respectively). The pre-2009 histogram is not well fit. However, it is not really constrained at the highest fluxes, which makes an interpretation tricky. The fit of the post-2009 data is worse than the fit to the whole data (red curve), but is still compatible with a Gaussian fit with similar parameters as the overall distribution.

The \textit{Swift} data is certainly influenced by data incompleteness, given that Swift is a pointing instrument and most observations are triggered during high flux states. While several coordinated campaigns took place over the last 20 years, these are also limited in duration compared to the full time range considered in this study. However, despite this issue, it is remarkable that the X-ray and B band distributions differ substantially, as both come from the same SED component and would thus be expected to show similar variability patterns. However, the optical flux drop between 2008 and 2009 is not mirrored in the X-ray data, which may explain parts of the discrepancy in the flux distributions.

The flux distributions also indicate the minimum and maximum fluxes attained; that is, the range of possible source fluxes. For the HE \g-rays, the 4\,d-binned fluxes range from $1.2\E{-8}\,$ph/cm$^2$/s to $4.5\E{-7}\,$ph/cm$^2$/s. Hence, the \g-ray fluxes are contained within a factor 40. The X-ray fluxes cover a range from $7.0\E{-12}\,$erg/cm$^2$/s to $5.5\E{-10}\,$erg/cm$^2$/s, which is roughly two orders of magnitude. 
The optical fluxes are harder to interpret. The total light curve covers fluxes from $2.5\E{-11}\,$erg/cm$^2$/s to $2.2\E{-10}\,$erg/cm$^2$/s, which is roughly an order of magnitude. However, the pre-2009 data starts at $9.3\E{-11}\,$erg/cm$^2$/s and reaches to the maximum value. Fluxes varied by less than a factor 3 in this subset. For post-2009 data, the starting flux is the global minimum, but the maximum is attained at $1.4\E{-10}\,$erg/cm$^2$/s -- roughly a factor 6 higher than the minimum.

The standard deviation (``std'') of the distributions clearly depends on energy. This is in-line with the attained flux ranges. Both measures indicate that higher-energy bands of a given SED component exhibit a broader range of variability than lower-energy bands. This is also true for the optical band, where the standard deviations of the separated time intervals agree within errors. Hence in this regard, there is no difference between the pre- and post-2009 intervals.

Besides data completeness, it should be noted that the flux distributions do not consider the statistical errors of each flux point. This is highlighted by the fact that neither the mean of the linear distribution nor that of the logarithmic distribution correspond to the (weighted) mean values given in Fig.~\ref{fig:2155_mwl_lc}, where the statistical errors are considered.

\subsection{Fractional variability} \label{sec:fvar}

\begin{figure*}[htb]
\centering
\includegraphics[width=0.98\textwidth]{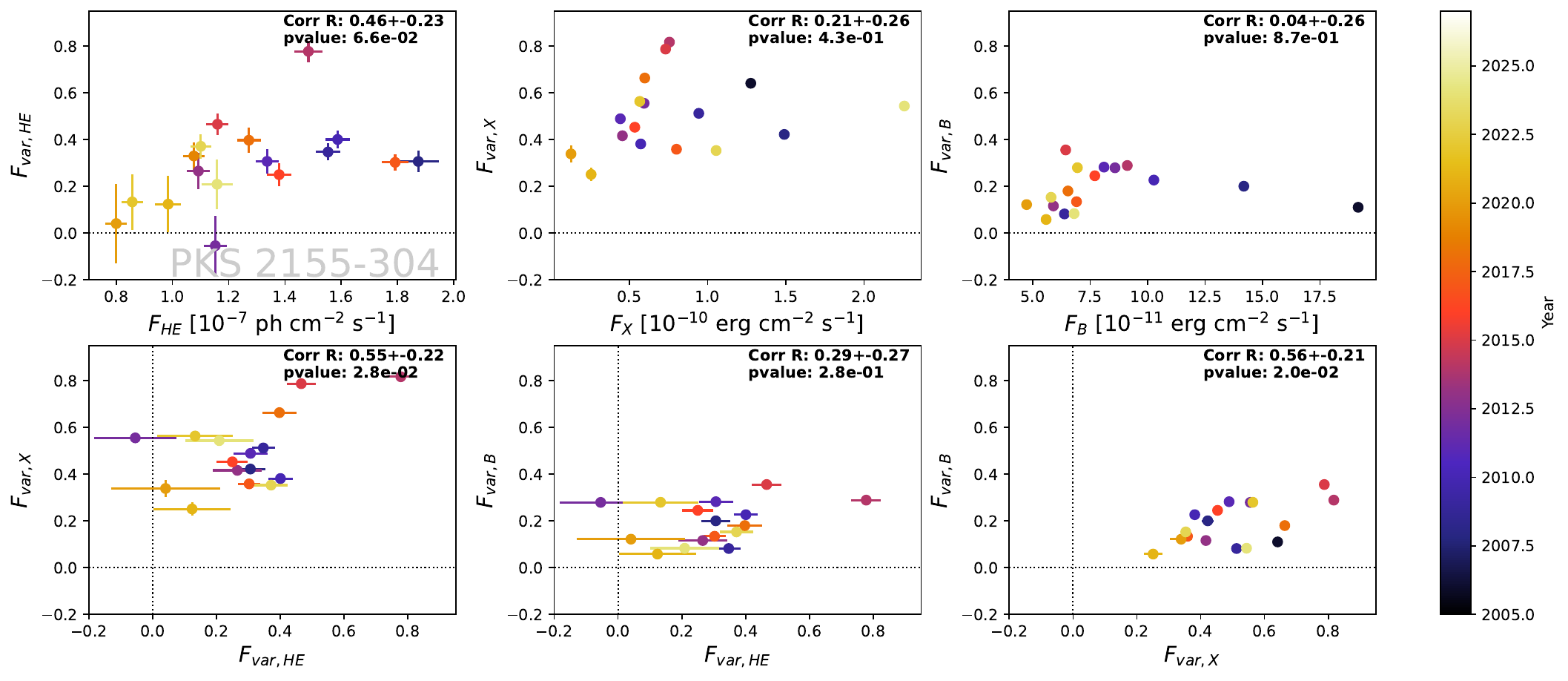}
\caption{(Top row) Year-wise \fvar\ (only if $\fvar\neq0$) as a function of yearly average flux for HE \g\ rays (left), X-ray (middle) and the B band (right). Horizontal error bars mark the average error in a given year. 
(Bottom row) Year-wise \fvar\ vs \fvar\ for X-rays versus HE \g\ rays (left), B band versus HE \g\ rays (middle) and B band versus X-rays (right).
In each panel, the Pearson R coefficient and p-value are given as in Fig.~\ref{fig:2155_flux_flux}. 
Grey dotted lines mark $\fvar=0$.
}
\label{fig:2155_fvar}
\end{figure*}

Figure~\ref{fig:2155_mwl_lc}(f) shows the fractional variability, \fvar, of each light curve for each calendar year of observation, while the horizontal dashed lines indicate the \fvar\ of the entire data set (and not the average from the year-wise values). The \fvar\ is calculated following \cite{vaughan+03}

\begin{align}
    \fvar = \sqrt{\frac{S^2-\overline{\sigma_{\rm err}^2}}{\overline{F}^2}}
    \label{eq:fvar}.
\end{align}
The associated uncertainty is \citep{poutanen+08}

\begin{align}
    \Delta F_{\rm var} = \sqrt{F_{\rm var}^2 + {\rm err}(\sigma_{NXS}^2)} - F_{\rm var}
    \label{eq:dfvar}
\end{align}
with the error on the normalised excess variance

\begin{align}
    {\rm err}(\sigma_{NXS}^2) = \sqrt{\frac{2 \overline{\sigma_{\rm err}^2}}{N\overline{F}^4} \left( 2F_{\rm var} \overline{F}^2 + \overline{\sigma_{\rm err}^2} \right)}
    \label{eq:errnormexcvar}.
\end{align}
In Eqs.~\eqref{eq:fvar} to~\eqref{eq:errnormexcvar}, $S^2$ and $\overline{F}$ are the variance and arithmetic mean of the light curve, $\overline{\sigma_{\rm err}^2}$ is the mean square of the measurement uncertainties, and $N$ is the number of data points in a given year. We only calculate $\fvar$ for $N>2$. 

In some cases, the mean square of the measurement uncertainties is larger than the variance, which implies a constant light curve. In such cases, we have derived the absolute value of the radicand in Eq.~\eqref{eq:fvar} and solved Eqs.~\eqref{eq:dfvar} and~\eqref{eq:errnormexcvar} with this, but afterwards defined \fvar\ as negative in order to highlight their different definition. 

Generally, it appears that the X-ray band exhibits the highest fractional variability, followed by the HE \g-rays and the optical band. Given the HBL nature of PKS~2155-304, this is not surprising. The main exceptions to this rule are 2010, 2014 and 2023, where the HE \g-ray fractional variability is equal to the X-ray one. In most cases, the evolution in \fvar\ seem to be relatively well correlated. Interestingly, the seemingly higher optical ``ground-level'' flux in 2005-2008 compared to later years does not influence this band's fractional variability.


In Fig.~\ref{fig:2155_fvar}, we study the correlations of the fractional variability with respect to flux and between each other. The correlations are again estimated with the Pearson R coefficient and the corresponding p value. 
With respect to fluxes -- which are the yearly averages -- none show a significant correlation, implying that the \fvar\ is independent of (average) flux state. Weighted constant fits are significantly ruled out (with at least $8\sigma$) in all three cases. 
%
None of the \fvar\ values are significantly correlated with each other, while a constant fit is ruled out in each case.

\section{X-ray spectral upturn} \label{sec:xupturn}



The X-ray spectral parameters, as shown in Fig.~\ref{fig:2155_mwl_lc}(d), are both variable. 
The average of the curvature is $0.21$ and both positive and negative values are detected.  
The latter case indicates concave spectral shape. While most of the occasions visible in Fig.~\ref{fig:2155_mwl_lc}(d) are not significant after careful checks (including compatibility with a simple power law), two subsequently taken observations stand out. These observations took place on April 28 and 29, 2012 (ObsIDs 00030795091 and 00030795092, respectively). The integrated flux level (in the energy range of  0.3-10\,keV) of  $(2.37 \pm 0.07) \times 10^{-11}\,\mathrm{erg\,cm^{-2}\,s^{-1}}$ is about half of the long-term average indicating a low state.

\begin{figure}[tb]
\centering
\includegraphics[width=0.48\textwidth]{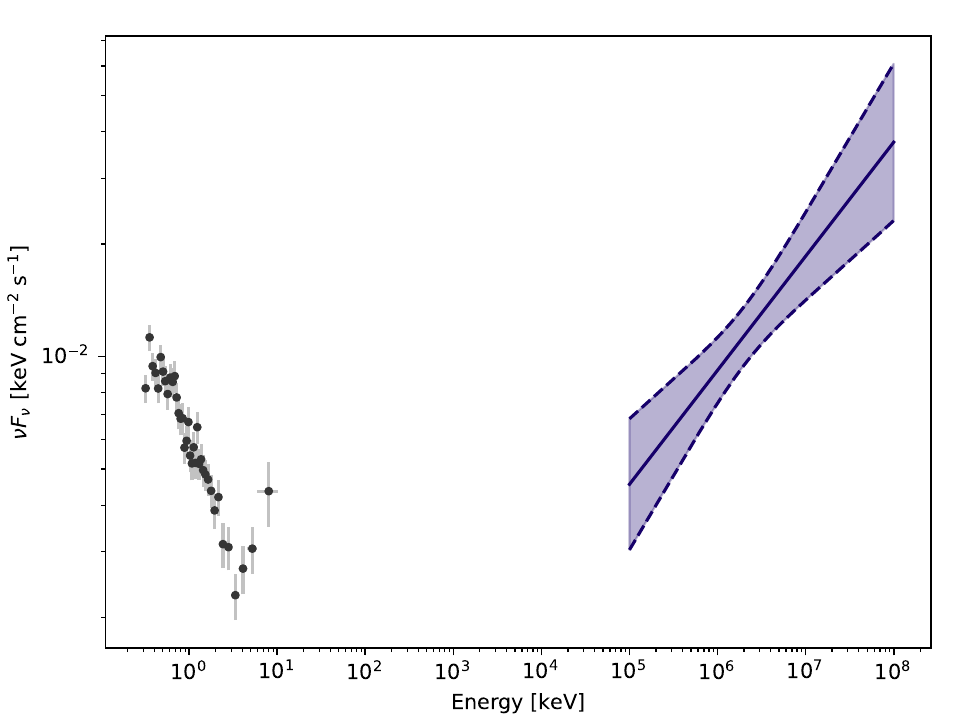}
\caption{Spectral energy distribution of PKS\,2155$-$304. X-ray spectrum represents the two merged Swift-XRT observations taken in April 2012, fitted with a broken power-law model. The bow-tie plotted in violet shows the quasi-simultaneous \fermi\ observations covering a period of 16\,d centred on April 28, 2012.
}
\label{fig:2155_upsed}
\end{figure}

After merging these two observations, a spectral fit with a broken-power-law model reveals a pronounced spectral upturn at an energy of $4.9\pm0.2\,$keV. 
The low-energy part of the spectrum is characterised by a photon index of $\Gamma_1 = 2.62\pm0.01$, which is only slightly softer than the long-term average, while the high-energy part is described by a photon index of $\Gamma_2 = 0.6\pm0.1$.
The resulting spectrum is shown in Fig.~\ref{fig:2155_upsed}.
The figure also shows as the violet bow-tie the HE \g-ray spectrum of \fermi\ for a quasi simultaneous period of observations. Namely, the \fermi\ data is integrated for a period of 16\,d centred on April 28, 2012. 
The LAT data are fitted with a single power-law model with photon index of $\Gamma=1.7\pm0.2$. The integrated flux above 100\,MeV is $(6.5 \pm 3.3) \times 10^{-8}~\mathrm{ph\,cm^{-2}\,s^{-1}}$, which is compatible with the long-term average.
Importantly, the X-ray upturn and the HE \g-ray bow-tie do not connect smoothly with each other.

Interestingly, the X-ray feature appears uniquely in our data set in April 2012. Even though similarly low flux states and soft spectra were observed with \textit{Swift} at other epochs, none of them indicates a similar upturn. 
However, both XMM-Newton and NuSTAR observations have revealed a break in the X-ray spectrum of PKS\,2155$-$304 during observations in 2006 \citep{zhang08pks2155} and 2013 \citep{madejski+16,hess+20pks2155}, respectively, with break energies at $\sim 4\,$keV and $\sim 10\,$keV. In the 2013 observations, the interpolation between the X-ray and HE \g-ray domains is much smoother than in our case. 
This strongly suggests that this feature is variable.
A more thorough investigation will be done elsewhere.

\section{Discussion} \label{sec:discus}

This paper presents 20-years of MWL observations of the blazar PKS\,2155$-$304, covering the period 2005 to 2024 and combining  $\gamma$-ray,  X-ray, and  optical-UV observations.
This long-term analysis provides a uniform view of the variability and spectral evolution of one of the brightest southern HBL-type blazars. 
The analysis presented in this work provides a long-term view of the source’s emission across different energy regimes.

The afore-described variability properties paint a complicated picture of PKS\,2155$-$304 and its evolution over these 20 years.
On the one hand, we see known or expected variability features. For instance, variability is more pronounced in higher energies of an SED component (here, the X rays) than at lower energies (optical and HE \g\ rays, respectively). This is also underlined by the fact that we find significant spectral variation in the X-ray domain following a harder-when-brighter trend, but none in the HE \g-ray band.
Flux-flux plots are suggestive of correlated activity between the various bands, and the flux distributions are compatible with log-normal distributions. The standard deviation of the log-normal fit function depends on energy suggesting a larger range of variations for higher energy bands of a given component -- in-line with the result from the \fvar\ study. 

On the other hand, several details do not follow common expectations.
A detailed cross-correlation study has not revealed any persistent correlation between the three bands over these 20 years. 
This is related to the flux-flux plot -- despite their suggestion of correlation -- not indicating any obvious (or singular) tracks. In other words, correlations vary from epoch to epoch and dilute the DCCF. As expected, the optical B and V bands are well correlated with the DCCF representing a near-auto correlation. Interestingly though, the DCCF is flat for up to $\tau=\pm 30\,$d with a meaningful drop only at $\tau=\pm70\,$d. This suggests that the optical flux keeps a long-lasting memory. This may be a consequence of the relatively slow and smooth variations, as suggested by the light curve in Fig.~\ref{fig:2155_mwl_lc}(e), where a given trend (flux rising or falling) seems to last for months.
The optical band suggests another oddity. Its ground-level apparently changed in 2009 dropping by a factor of a few. However, the statistical properties (\fvar\ and standard deviation of the flux distribution) were seemingly not influenced by this change.
Details of the fractional variability study also show an unexpected result: The value of the \fvar\ is not correlated with the integrated flux level. For instance, the maximum \fvar\ does not necessarily correspond to the highest average flux.

While the X-ray band, as stated, does show harder-when-brighter behaviour, it is worth pointing out that this trend is not uniform. We observe the slope of the correlation differing from year to year, which means that the hardness of the X-ray spectrum does not directly correlate with the flux level. 
This is different from other HBL-type sources, like Mrk\,421 and Mrk\,501, where the index-vs-flux plot shows a relatively narrow (even though not linear) relation \citep[e.g.,][]{Taylor+26}. 
Hence, there does not seem to be an universal cause for the flux and spectral changes in PKS\,2155$-$304.
In this regard, it worth noting that its X-ray photon index never drops below $2.0$ (the single instance, where it happened, is not significant), even though the index gets close to $2.0$ occasionally. Such hard spectra are attributed to so-called extreme HBLs \citep[e.g.,][]{goswami+24}. While other HBL sources 
can exhibit such extreme states from time to time \citep[e.g.,][]{ahnen+18_mrk501,abeysekara20_mrk421,acciari+20_2344,acciari20_1959}, PKS\,2155$-$304 apparently does not do this even during very bright flares like the one in 2024.

The HE \g-ray band does not show any relation between integrated flux and spectral index. However, we have found several cases where the index increases significantly to values above 2. This indicates a soft HE \g-ray spectrum, which is unusual for an HBL, where the peak of the \g-ray SED component is typically located at $\sim100\,$GeV. A soft HE spectrum indicates that the peak position must have dropped to lower energies; perhaps even below $100\,$MeV. We do not find any relation with flux nor a specific clustering in time for these features, which happened in about 3\% of our time bins. There is also no relation with the \textit{Swift} data, even though the low number of occurrences along with the \textit{Swift} cadence results in only 7 contemporaneous bins. A more detailed study of these events requires contemporaneous VHE \g-ray data in order to properly assess the state of the high-energy SED component. This is beyond the scope of this paper and will be pursued elsewhere.

These observations appear difficult to be reconciled within a homogeneous one-zone model, which is still the go-to model for most modelling attempts. The lack of (long-term) correlations, as well as the various flux and variation properties suggest that the underlying source parameters (particle distribution, magnetic field, emission region size) vary in a complicated manner. 
While it may be possible to model this with a time-dependent one-zone code -- which is beyond the scope of this paper -- the required almost chaotic changes are more suggestive of turbulent and/or multi-zone approaches \citep[e.g.,][]{marscher14}. 
A more elaborate discussion is presented in \cite{ZW26}.

The temporal variations of the X-ray spectral parameters revealed a special event in our data set, namely the significantly concave X-ray spectrum in April 2012. This is the third instance, such an upturn has been observed in the X-ray spectrum of PKS\,2155$-$304 \citep{zhang08pks2155,madejski+16}.
While an X-ray spectral upturn is normally observed in IBL-type blazars, it is not a common feature in HBL-type ones \citep[see, e.g.,][]{2016MNRAS.458...56W}. It has only been reported in a few HBLs so far, including Mrk\,421 \citep{Kataoka_421}, 5BZB\,J0630$-$2406 \citep[e.g.][]{Zaballa_0630}, and 1ES\,0229+200 \citep{Wierzcholska_0229}. In all cases, these features were detected in a low X-ray state. 

This spectral feature suggests the transition between two distinct emission components.
In this scenario, the steeply falling synchrotron tail from the highest-energy electrons meets the rising segment of a different component formed by inverse-Compton or hadronically-induced emission processes. 
In the latter case, the excess hard X-ray emission may trace the Bethe–Heitler or \g\g\ pair-production processes, in which relativistic protons interact with ambient photons, producing secondary electron–positron pairs that radiate synchrotron emission in the hard X-ray band \citep{PetropoulouBH}.
Such an interpretation is consistent with predictions of lepto-hadronic models, where the secondary synchrotron component emerges as a broad hard excess between the primary electron-synchrotron and \g-ray components \citep[e.g.,][]{goswami+24,Chandra+25}, where the latter may be inverse-Compton or proton-synchrotron emission.
A third option is a multi-zone model, where the additional component may reveal a spatially separate emission zone. While these separate emission regions reveal themselves in different ways, they have been strongly suggested for a couple of different sources and blazar types \citep[e.g.,][]{zachariaswagner16,hess20,hess23}. 

Unlike in other cases of this upturn in PKS\,2155$-$304 \citep[e.g.][]{hess+20pks2155}, the interpolation between the X-ray and the HE \g-ray spectrum in the case uncovered here (see Fig.~\ref{fig:2155_upsed}) is not smooth. This suggests that the radiative component responsible for this upturn stems from a separate process than the \g-ray-producing one. Furthermore, our data in combination with the works by other authors \citep{zhang08pks2155,madejski+16} indicate that this feature is intermittent; both in terms of its flux and its location in energy space. This makes it less likely to be connected to the inverse-Compton component within a one-zone model, as it should vary similarly as (or even less than) the \g-ray flux. A hadronic scenario or a multi-zone interpretation thus appear more plausible. Observing this feature with an X-ray polarimeter would help to narrow down its radiative nature.

\subsection{Data completeness} \label{sec:complete}

The long-term data set used in this work inevitably suffers from non-uniform temporal coverage, particularly with the \textit{Swift} observations.
While \textit{Fermi}-LAT has provided quasi-continuous all-sky monitoring since 2008, the \textit{Swift}-XRT and \textit{Swift}-UVOT data are limited by the satellite’s pointing schedule and were often obtained during high-activity states or coordinated MWL campaigns.
This introduces an intrinsic bias toward elevated flux levels and underrepresents quiescent intervals, especially in the X-ray and optical bands. 
As a consequence, flux distributions derived from \textit{Swift} data are affected by incompleteness, which can artificially enhance the apparent variability amplitude and distort statistical properties such as log-normality.
The irregular sampling also weakens the robustness of cross-correlation analyses, as large temporal gaps reduce sensitivity to potential interband lags on timescales of days to weeks.

Similarly, year-by-year estimates of the fractional variability (\fvar) can be inflated in sparsely sampled periods, or even rendered unreliable when only a few observations are available \citep[see also the detailed discussion in][]{schleicher+19}. 
This effect is evident in several years where the \fvar\ is poorly constrained despite moderate flux changes. 
The lack of continuous coverage may also obscure short-term flares or transitional states, leading to an incomplete picture of the temporal and spectral variability. 

Future monitoring with more homogeneous cadence in several bands, ideally through coordinated multi-instrument campaigns, would mitigate these effects and allow for a more accurate quantification of the variability and its temporal correlations.

\section{Summary} \label{sec:sum}

We have presented a 20-year MWL variability study of the blazar PKS\,2155$-$304, using \textit{Fermi}-LAT  data and \textit{Swift} XRT and  UVOT observations obtained between 2005 and 2024.
The main results of this work are summarised as follows:

\begin{itemize}
    \item The long-term data sets reveal pronounced flux variability in the temporal and spectral domain across all studied energy bands.
    \item The flux distributions are well described by log-normal functions. The optical flux exhibits an odd baseline drop happening in 2009.
    \item Significant flux-flux correlations are observed between all bands. However, the discrete cross-correlation analysis reveals no persistent interband lags, suggesting that the correlations vary between epochs.
    \item The X-ray band exhibits a consistent harder-when-brighter trend, although the slope of this relation changes with time, implying non-universal causes for the flux and spectral changes.
    \item The $\gamma$-ray spectra show only weak spectral variability, with no clear dependence of photon index on flux level. In about 3\% of the flux bins, the \g-ray spectrum is soft with an index significantly higher than 2. This suggests a shift of the SED peak position from the nominal $\sim 100\,$GeV to $\lesssim100\,$MeV, a potentially remarkable feature for an HBL.
    \item The fractional variability, \fvar, is highest in X rays, moderate in $\gamma$ rays, and lowest in the optical band, consistent with expectations for an HBL. The fractional variability does not correlate with the average flux level, indicating that strong variability does not necessarily coincide with high-average-flux states.
    \item A concave X-ray spectrum was detected in the low-flux \textit{Swift}-XRT observations from April~2012, well described by a broken power-law model with a spectral upturn at $\sim5\,$keV. This spectral upturn may indicate either the transition between the synchrotron tail and the rising inverse-Compton  or it may be produced by Bethe-Heitler or \g\g\ pair-induced synchrotron emission, implying the presence of highly energetic protons and a possible hadronic contribution to the jet emission. A third option is a spatially-separated component becoming visible at these states. Interestingly, the feature itself is variable, as it was not detected during similarly low X-ray states in our data set.
    \item The overall variability behaviour, including the absence of persistent cross-band correlations and the non-uniform spectral evolution, cannot be fully explained within a simple homogeneous one-zone model, favouring instead a multi-zone or turbulent jet scenario.
\end{itemize}

Finally, we underline here that although the \textit{Swift} observations suffer from non-uniform temporal sampling, the combined 20-year data set provides a unique view of the long-term variability of PKS\,2155$-$304. 
This study demonstrates the diagnostic power of long-term, multi-wavelength monitoring to help understanding the complex temporal and spectral behaviour of blazars. Continuous and coordinated observations over extended timescales remain crucial for disentangling the physical mechanisms driving jet emission and for assessing the potential contribution of blazars to the cosmic-ray and neutrino populations.

%
%
\section{Acknowledgement}
The authors thank the anonymous referees for constructive comments that helped to improve the manuscript.
The authors acknowledge stimulating discussions with G.L.~Taylor and S.~Wagner regarding the variability aspects of blazars.
The authors gratefully acknowledge the Polish high-performance computing infrastructure PLGrid (HPC Center: ACK Cyfronet AGH) for providing computer facilities and support within computational grant no. \text{PLG/2024/017925}. %

\section{Author contributions}
A.~Wierzcholska (Conceptualization, Data curation, Formal analysis, Writing – original draft),
M.~Zacharias (Conceptualization, Methodology, Writing – original draft)

\section{Funding sources}
The project is co-financed by the Polish National Agency for Academic Exchange. %
The project on which this report is based was funded by the Bundesministerium für Bildung und Forschung (BMBF, Ministry of Education and Research). Responsibility for the content of this publication lies with the author. 
This work is funded in parts by the Deutsche Forschungsgemeinschaft (DFG, German Research Foundation) -- project number 460248186 (PUNCH4NFDI). %
%

%
%

%

\bibliographystyle{elsarticle-harv} 
\bibliography{references}
\end{document}